\documentclass[prl,twocolumn]{revtex4-1}

\usepackage{float}
\usepackage{amsmath}
\usepackage{amsfonts}
\usepackage{graphicx}
\usepackage{parskip}
\usepackage{algorithmicx}
\usepackage{algpseudocode}
\usepackage{algorithm}
\usepackage{circuitikz}
\usepackage{listings}
\usepackage{color}

\newcommand{\mytensor}[1]{\mathbf{#1}}
\newcommand{\p}[2]{\frac{\partial #1}{\partial #2}}
\newcommand{\tr}[0]{\mathrm{tr}}

\begin{document}
\begin{abstract}
  We propose a model for the evolution of the conductivity tensor for a
  flowing suspension of electrically conductive particles.
  We use discrete particle numerical simulations together with a continuum physical framework
  to construct an evolution law for the suspension microsutructure during flow.
  This model is then coupled with a relationship between the microstructure and the electrical conductivity tensor.
  The parameters of the joint model are fit experimentally using rheo-electrical conductivity measurements of
  carbon black suspensions under flow over a range of shear rates.
  The model is applied to the case of steady shearing as well as time-varying conductivity of unsteady flow experiments.
  We find that the model prediction agrees closely with the measured experimental data in all cases.
\end{abstract}

\title{
  Coupled dynamics of flow, microstructure, and conductivity in sheared suspensions
}
\author{Tyler Olsen}\affiliation{Massachusetts Institute of Technology, Department of Mechanical Engineering}
\author{Ahmed Helal}\affiliation{Massachusetts Institute of Technology, Department of Mechanical Engineering}
\author{Gareth McKinley}\affiliation{Massachusetts Institute of Technology, Department of Mechanical Engineering}
\author{Ken Kamrin}\email{kkamrin@mit.edu}\affiliation{Massachusetts Institute of Technology, Department of Mechanical Engineering}
\maketitle


\textit{Introduction:}
Microstructural anisotropy has been an active area of research for decades.
It plays a critical role in biomechanics \cite{Bischoff2002,Deneweth2013},
plasticity \cite{Frederick2007}, granular materials
\cite{Azema2012,DaCruz2005,Mehrabadi1982,Oda1982,Radjai2012,Satake1978,Sun2011}, 
liquid crystals \cite{Stephen1974}, and more.
Some materials, such as elastic composites, have fixed anisotropy that does not evolve over time.
However, other materials may develop anisotropy due to deformation, e.g. kinematic hardening of 
solids \cite{Frederick2007}, or due to an externally-applied field, such as an electric field, 
as is typical of liquid crystals \cite{Stephen1974}.

Of particular interest in this study is the flow-induced anisotropy of colloidal suspensions
\cite{Grenard2011,Grenard2014,Koumakis2012,Mohraz2005,Negi2009,Osuji2008,Osuji2008a,Park2013,Santos2013,Trappe2001,Trappe2000}.
Suspensions of carbon black, an electrically-conductive form of carbon, have recently
found application in a class of semi-solid batteries called ``flow batteries'' \cite{Duduta2011,Youssry2013}.
At concentrations above the percolation threshold, the carbon black creates an electrically 
conductive network inside the flowing electrolytes
of the battery, allowing for higher reaction rates and overall system efficiency.
However, it has been experimentally demonstrated that the networks in these carbon 
suspensions are highly sensitive to shearing \cite{Alig2007,Amari1990,Bauhofer2010,Schulz2010}.
In these studies, the conductivity of the carbon network drops precipitously with shear
and recovers dynamically when brought to rest.
This has serious implications for battery performance if the evolution of
network structure and conductivity are not properly considered during design.
Recent studies \cite{Smith2014} on optimizing the efficiency of a flow battery 
have neglected the effect of a shear-induced drop in suspension conductivity.
In addition to the drop in conductivity, it has been observed that the suspension microstructure
becomes anisotropic during shearing flow, which can lead to anisotropic conductivity
\cite{Hoekstra2003,Morris2002,Vermant2005}.
In this study, we use discrete-particle simulations and continuum physical arguments 
to derive a general constitutive law for the flow-induced evolution
of a tensor-valued measure of suspension network anisotropy.  
We couple this with a nonlinear structure-conductivity relation, and show that the 
calibrated joint model makes quantitative predictions of conductivity evolution in 
many different experimental flows of carbon black.

We use a \textit{fabric tensor} 
to describe the structure of the particle network in suspension.
This concept was originally devised to describe the contact network in granular materials
\cite{Mehrabadi1982,Oda1982,Satake1978}.
The fabric tensor  can be defined at the particle level
with the relation
\begin{equation}
  \label{eq:ParticleFabric}
  \mytensor{A}^P = \sum_{i=1}^{N_{contacts}} \mytensor{n}^{(i)} \otimes \mytensor{n}^{(i)}
\end{equation}
where $\otimes$ denotes the dyadic product and $\mytensor{n}^{(i)}$ the contact unit normals.
It is often illustrative to examine the average fabric of a group
of particles rather than the particle-level information, i.e. $\mytensor{A}=\langle  \mytensor{A}^P \rangle_P$.
This definition yields a number of useful properties. 
The trace of $\mytensor{A}^P$ is equal to the coordination number of contacts on a particle. 
Consequently, $\tr\mytensor{A}$ represents the average coordination number, $Z$, of a group of particles. 
Second, this definition results in a symmetric, positive semi-definite tensor. 
This is appealing, because these tensorial properties are shared by the conductivity tensor $\mytensor{K}$.

In a previous numerical study \cite{Olsen2015}, we modeled the conductivity tensor of a particle network 
as a function of the average fabric tensor $\mytensor{A}$, by assuming the network could be represented 
by a regular  lattice of identical particles with the same average fabric tensor. 
The fabric-lattice relationship can be inverted to obtain a model for conductivity as shown below in 3D:
\begin{equation}
  \label{eq:KAtensor3D}
  \mytensor{K} = k_1 \frac{\left(\tr\mytensor{A} - 2\right)^2}{\det\mytensor{A}} \mytensor{A}.
\end{equation}
The above was validated for computer-generated random sphere networks, but was never tested experimentally; 
a byproduct of its usage in the current study is a de facto experimental test and a check on its robustness for non-spherical particles.

\textit{General Evolution Law:}
We set out to develop a continuum model to accurately characterize the evolution of flowing particle 
networks indicated by the aforementioned experiments.
Although the fundamental quantity---the particle network---is composed of discrete units,
we make a continuum approximation such that quantities at a point represent local spatial averages, e.g. velocity or fabric.
This is a valid approximation since typical applications of these particle networks are several
orders of magnitude larger than the constituents of the networks.

We define the velocity gradient $\mytensor{L} = \p{\mytensor{v}}{\mytensor{x}}$,
the strain-rate tensor $\mytensor{D} = \frac{1}{2}\left( \mytensor{L} + \mytensor{L}^T   \right)$,
and the spin tensor $\mytensor{W} = \frac{1}{2}\left( \mytensor{L} - \mytensor{L}^T   \right)$.
We postulate a fabric evolution law of the form 
$\dot{\mytensor{A}} = \boldsymbol{\Psi}(\mytensor{A}, \mytensor{L})$,
where $\dot{\mytensor{A}}$ denotes the material time derivative of $\mytensor{A}$.
In order for an evolution law such as this to be indifferent under a change in an observer's frame
of reference, the evolution law must be expressible as
$\dot{\mytensor{A}} = \mytensor{WA} - \mytensor{AW} + \hat{\boldsymbol{\Psi}}(\mytensor{A}, \mytensor{D})$,
where  $\hat{\boldsymbol{\Psi}}$ is an isotropic function of the fabric tensor and the stretching tensor \cite{Hand1961a}. 
A representation theorem for isotropic functions of 3$\times$3 symmetric tensors \cite{Rivlin1955} can be applied, allowing us to write
\begin{align}
  \label{eq:EvolutionGeneral}
  \dot{\mytensor{A}} + \mytensor{AW} - &\mytensor{WA} = 
  c_1 \mytensor{1} + c_2 \mytensor{A} + c_3 \mytensor{D} +
  c_4 \mytensor{A}^2 + c_5 \mytensor{D}^2 \nonumber\\
  + &c_6 (\mytensor{AD} + \mytensor{DA}) +
  c_7 (\mytensor{A}^2\mytensor{D} + \mytensor{D}\mytensor{A}^2) \nonumber\\
  + &c_8 (\mytensor{A}\mytensor{D}^2 + \mytensor{D}^2\mytensor{A}) + 
  c_9 (\mytensor{A}^2\mytensor{D}^2 + \mytensor{D}^2\mytensor{A}^2)
\end{align}
In the above expression, $c_i = c_i(\mathcal{I}_{\mytensor{A},\mytensor{D}})$,
where the full set of simultaneous invariants of $\mytensor{A}$ and $\mytensor{D}$ is 
$\mathcal{I}_{\mytensor{A},\mytensor{D}}=\left( \cup_{\alpha,\beta<3}\, \tr\mytensor{A}^{\alpha}\mytensor{D}^{\beta}\right)\cup \tr\mytensor{A}^3 \cup\tr\mytensor{D}^3$.
The left-hand side of \eqref{eq:EvolutionGeneral} is the co-rotational time derivative
of $\mytensor{A}$, or Jaumann rate,  given the symbol $\mathring{\mytensor{A}}$.
In general, the left-hand side can be any objective time derivative of the tensor field;  
all specializations atop the Lie derivative \cite{Marsden1994}.
Without loss of generality,  we chose to use the co-rotational rate of $\mytensor{A}$ for ease of modeling; 
other objective rates, such as the contravariant or covariant time derivatives, 
do not equal $\dot{\mathbf{A}}$ in a spin-free flow \cite{Gurtin2010}.

The general evolution law in \eqref{eq:EvolutionGeneral} has a large
number of scalar functions that must be specified.  
For simplicity, we neglect higher-order tensorial terms by setting 
$c_i(\mathcal{I}_{\mytensor{A},\mytensor{D}}) \equiv 0$ for $i \ge 4$.
This leaves the quasi-linear form
\begin{equation}
  \label{eq:EvolutionReduced}
  \mathring{\mytensor{A}} = c_1 \mytensor{1} + c_2 \mytensor{A} + c_3 \mytensor{D}.
\end{equation}
The task of modeling, therefore, is reduced to choosing physically meaningful functions for $c_1$, $c_2$, and $c_3$.

By examining the effect of each term on the evolution of the fabric, 
some physical constraints must be satisfied by the choice of the $c_i$.
First, the fabric will be positive, isotropic, and unchanging after a long 
period of no flow; anisotropy induced by flow must relax away over time. 
This implies
\begin{equation}
  \label{eq:signConstraintC1}
  c_1(\mathcal{I}_{\mytensor{A},\mytensor{D}}) > 0; \quad
  c_2(\mathcal{I}_{\mytensor{A},\mytensor{D}}) < 0 \quad \forall \;\mytensor{A},\mytensor{D}.
\end{equation}
If either of these constraints are violated, then the fabric would either decay 
away to a non-positive isotropic state or diverge.

Second, contacts are formed on the compressive axis of shearing flow and broken on 
the extension axis (as experimentally confirmed in Hoekstra et al.\cite{Hoekstra2003}).
This gives us the condition
\begin{equation}
  \label{eq:signConstraintC3}
  c_3(\mathcal{I}_{\mytensor{A},\mytensor{D}}) < 0 \quad \forall \; \mytensor{A},\mytensor{D}.
\end{equation}

Third, while the electrical conductivity decreases with increasing shear rate, 
the conductivity never reaches zero despite the fluid being a strong insulator \cite{Amari1990}.
Based on the conductivity model assumption in \eqref{eq:KAtensor3D}, this implies
that $\tr\mytensor{A}$ remains above 2 at all times.
This condition implies
\begin{equation}
  \label{eq:traceConstraint}
  -\frac{c_1}{c_2} > \frac{2}{3} \quad \forall \;\mytensor{A},\mytensor{D}.
\end{equation}

Finally, as the fabric is necessarily positive semi-definite, the evolution law must guarantee this property is preserved.
A sufficient condition for this, as derived in the Supplemental Materials, is
\begin{equation}
  \label{eq:c1c3Constraint}
  \frac{c_1}{c_3} \leq -\sqrt{\frac{2}{3}}\,|\mytensor{D}|.
\end{equation}

\textit{Numerical Experiments:}
To gain insight on the relaxation behavior of the fabric, characterized 
through $c_1$ and $c_2$, we created  a discrete particle aggregation code.
In the code, 100,000 particles are seeded into a periodic box at a 1.5\% 
volume fraction and allowed to diffuse.
Particles and clusters are assigned velocities such that the distance that a cluster moves
in a single time step is drawn from a Gaussian distribution with variance $D\Delta t$, where
$D$ is the diffusion coefficient, and $\Delta t$ is the simulation time step length.
As hit-and-stick behavior is typical in diffusion-limited
aggregation \cite{Jr1981}, all clusters in contact after a step are deemed to stick, 
creating a larger cluster.
As clusters grow, the diffusion coefficient is adjusted according to $D=D_0/N$, where $D_0$
is the diffusion coefficient for a single particle, and $N$ is the number of particles 
in a cluster.  
Because clusters typically have a snake-like shape, dominated by long strands of particles, 
this relation was chosen for its relation to the asymptotic solution for the diffusion 
coefficient of a string of $N$ particles \cite{Happel2012}.

\begin{figure}
  \centering
  \includegraphics[width=3.25in,trim=.55in 0.15in .45in .2in, clip]{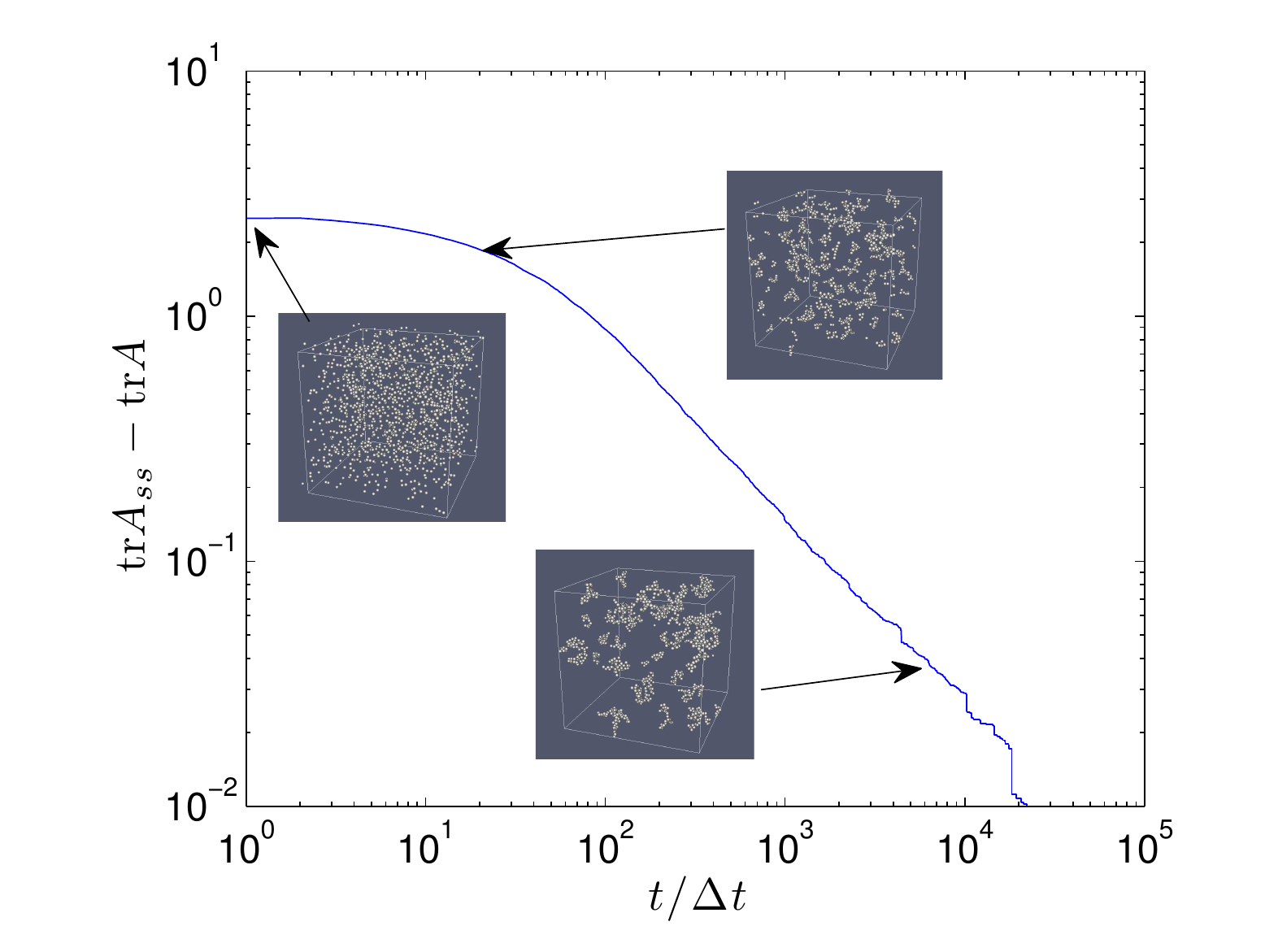}
  \caption{
    Log-log plot of $\tr\mytensor{A}_{ss}-\tr\mytensor{A}$ vs time shows power-law nature
    of $\tr\mytensor{A}$ to steady state.
    Inset images show a typical particle configure at a given point on the curve.
  }
  \label{fig:trAvsTime}
\end{figure}

The simulation results are shown in Fig \ref{fig:trAvsTime}, where 
$\mytensor{A}_{ss}$ is the steady-state fabric.  
After a brief startup period, the deviation from steady state 
$(\tr\mytensor{A}_{ss}-\tr\mytensor{A})$ and 
time relate through a power law
\begin{equation}
  \label{eq:simulationFit}
  (\tr\mytensor{A}_{ss} - \tr\mytensor{A}) \sim (t/\Delta t)^{-0.745}
\end{equation}

\textit{Form of evolution coefficients:}
Motivated by the power-law decay of $\tr\mytensor{A}$ to steady state we have just determined,
we choose the following functional forms for the $c_i$ coefficients:
\begin{align}
  \label{eq:c1Function}
  c_1 &= \frac{1}{3} \left(\frac{Z_0}{\tau}\left(Z_0 - \tr\mytensor{A}\right)^n 
  + \beta Z_\infty \left|\mytensor{D}\right| \right)\\
  \label{eq:c2Function}
  c_2 &= -\left(\frac{1}{\tau}\left(Z_0 - \tr\mytensor{A}\right)^n + 
  \beta \left|\mytensor{D}\right| \right)\\
  \label{eq:c3Function}
  c_3 &= \alpha
\end{align}
where $Z_0 = \tr\mytensor{A}_{ss}$ (in the absence of flow),
$Z_\infty = \tr\mytensor{A}_{ss}$ as $\left|\mytensor{D}\right|\rightarrow \infty$,
$\tau$ is the time scale of thermal fabric relaxation,
$\beta$ reflects the network creation or disruption due
to non-affine flow perturbations,
$\alpha$ is the initial rate of anisotropy formation
when started from an isotropic state, and $n$ characterizes the relaxation
to the no-flow steady state (see Fig \ref{fig:trAvsTime}).
The above constraints imply the following inequalities on these parameters:
$Z_0,Z_\infty > 2$;
$\tau,\beta > 0 > \sqrt{2/3}\, \alpha > -\beta Z_\infty/3$.
The value of $n$ can be predicted from the discrete simulation data by 
taking the trace of Eq \eqref{eq:EvolutionReduced}, setting $\mytensor{L}$
to $\mytensor{0}$, 
integrating to find $\tr\mytensor{A}$ as a function of time, and relating the answer 
back to the power-law in Eq \ref{eq:simulationFit}. 
Using this, we find $n=1.34$.

\begin{figure}
  \centering
  \includegraphics[width=3.3in]{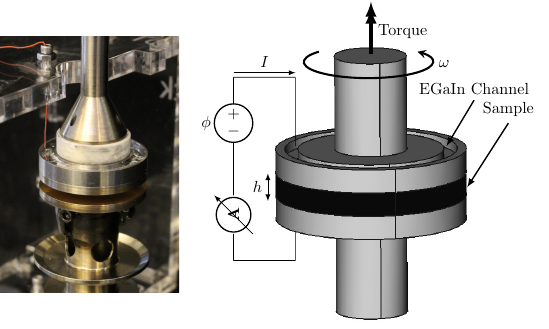}
  \caption{Device used to perform rheo-electric measurements.}
  \label{fig:expSetup}
\end{figure}

\textit{Experimental Methods:}
The system studied is a carbon black suspension prepared in the absense of dispersant
by mixing carbon black particles in a light mineral oil.
Details of materials and preparation can be found in the supplementary material.
Simultaneous rheo-electric measurements were performed using a custom device,
described in more detail in the supplementary material.
A schematic of the device used to perform the rheo-electric measurements is shown in Fig
\ref{fig:expSetup}.

Two sets of experiments were performed using the setup described above.
The first was a set of steady-state current measurements taken at nominal shear rates
in the range $\dot\Gamma=\omega R/h \in \left[0, 300 \right]s^{-1}$.  
At each shear rate, both the current and stress were allowed to equilibrate before
the measurement was recorded to ensure that it had relaxed to its steady value.
The shear rates were swept in descending order to mitigate complications such as shear-induced 
phase separation that arise at low shear rates, below $\dot\Gamma \sim 20s^{-1}$ \cite{Helal201x}.

The second dataset was a collection of transient ramp tests wherein current data was
collected continuously for the duration of the test.
The ramp tests consisted of 5 minutes of nominal shear rate $\dot{\Gamma}=\dot{\Gamma}_1$,
ramping linearly to $\dot\Gamma_2$ over duration $t_R$, holding for 5 minutes,
and abruptly setting $\dot\Gamma = 0$, collecting data for 15 additional minutes.
Pre-shear at $\dot\Gamma =100s^{-1}$ was applied for 5 minutes before each test
to ensure consistent initial conditions.

The parameters $Z_0$, $Z_\infty$, $\beta$, and $\alpha$ were fitted to the steady-state 
current measurements using the $\mathtt{fminunc}$ Matlab optimization routine, 
minimizing the squared difference between predicted and measured electrical currents.
The $\tau$ parameter, which is primarily responsible for controlling the fabric's relaxation time, 
was chosen from the experimental dynamics of the ramp tests.
Lastly, the $k_1$ parameter can be chosen to match the steady-state current observed at $\dot\Gamma= 0$.

The predicted current is calculated by evaluating the integral
\begin{equation}
  \label{eq:CurrentIntegral}
  I(t) = 2\pi \int_0^R K_{zz}(\mytensor{A}(r,t))\cdot\frac{\phi}{h} r\,dr
\end{equation}
where $K_{zz}$ is the component of conductivity perpendicular to the plate, $\phi$ is the
applied potential difference across the plates, $h$ is the plate separation,
and $R$ is the plate radius.
Note that the fabric tensor is a function of radial position;
each point along a radius is subjected to a different shear rate due to the applied 
torsional motion, and thus evolves differently.

The form of \eqref{eq:CurrentIntegral} can be modified slightly in order to solve
for the current at steady-state for a given nominal shear rate $\dot\Gamma$, 
by substituting $\mytensor{A}(r,t)$ with $ \mytensor{A}_{ss}( \dot\gamma= \dot\Gamma r/R )$
where $\mytensor{A}_{ss}(\dot{\gamma})$ is the steady-state fabric tensor, 
whose components are obtained algebraically from Eq \ref{eq:EvolutionReduced} by defining 
$\mytensor{D}$ and $\mytensor{W}$ to correspond to simple shearing at $\dot{\gamma}$ 
and setting $\dot{\mytensor{A}}=\mytensor{0}$.
A detailed outline of the fitting procedure can be found in the supplementary material.

To simulate the electrical current for a transient test, the evolution law
must be directly integrated at each point across the disk where the conductivity will be evaluated.
To do this, one must input the time-dependent nominal shear-rate $\dot\Gamma(t)$ from the experimental protocol.
The evolution law was numerically integrated using the 4th-order Runge-Kutta method implemented
in the Matlab function $\mathtt{ode45}$.
After the fabric is known for all time at each point, \eqref{eq:CurrentIntegral}
can be approximated directly using a Riemann sum to give the predicted current
as a function of time.


\begin{figure}
  \centering
  \includegraphics[width=3in]{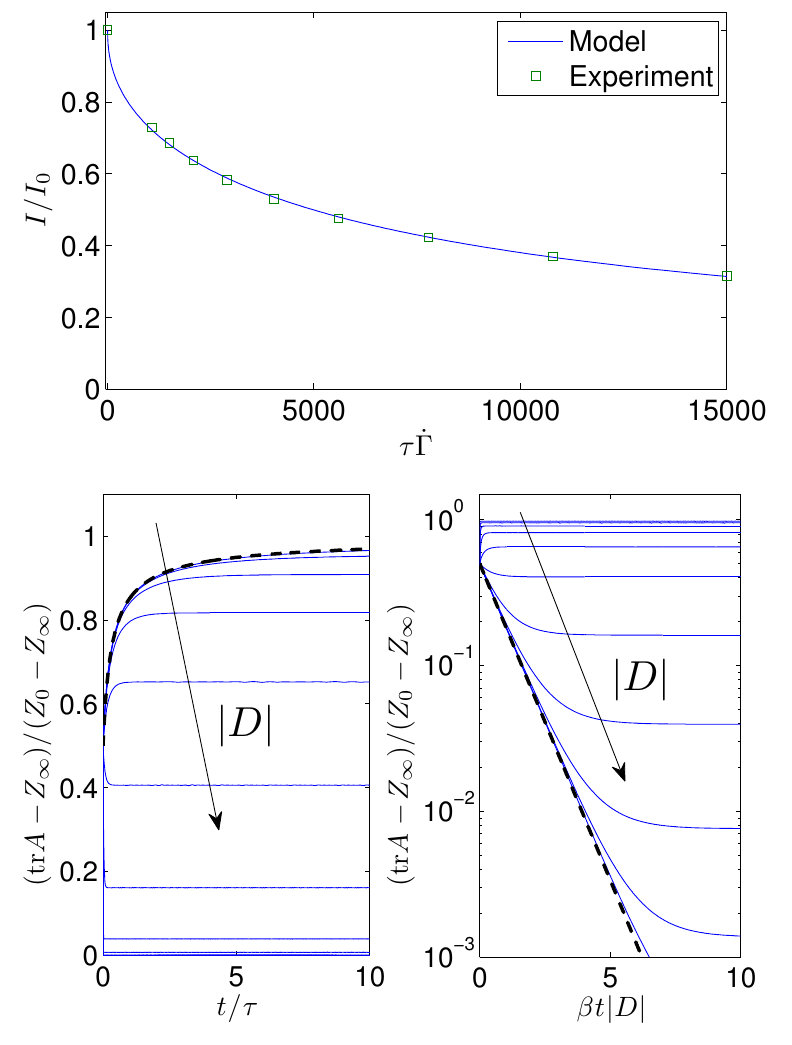}
  \caption{
    Top:
    Model fit of steady-state current measurements at different nominal shear rates.
    Values are normalized by static current measurement $I_0$.
    Bottom: Evolution of $Z=\tr\mytensor{A}$ in pure shearing under different shear rates, as predicted by the model.
    (Left) Small $\left|\mytensor{D}\right|$ yields power-law approach of $Z$
    to steady state; $|Z-Z_0| \sim t^{-1/n}$ as $|D|\to 0$, dotted line.
    (Right) Large $\left|\mytensor{D}\right|$ yields exponential approach to steady state;
    $\log(Z -Z_{\infty})\sim- \left|\mytensor{D}\right| t$ as $|D|\to \infty$, dotted line.
  }
  \label{fig:2:NLSSModelfit}
\end{figure}

\begin{figure}
  \centering
  \includegraphics[width=3in,trim=.0in .2in .0in .1in, clip]{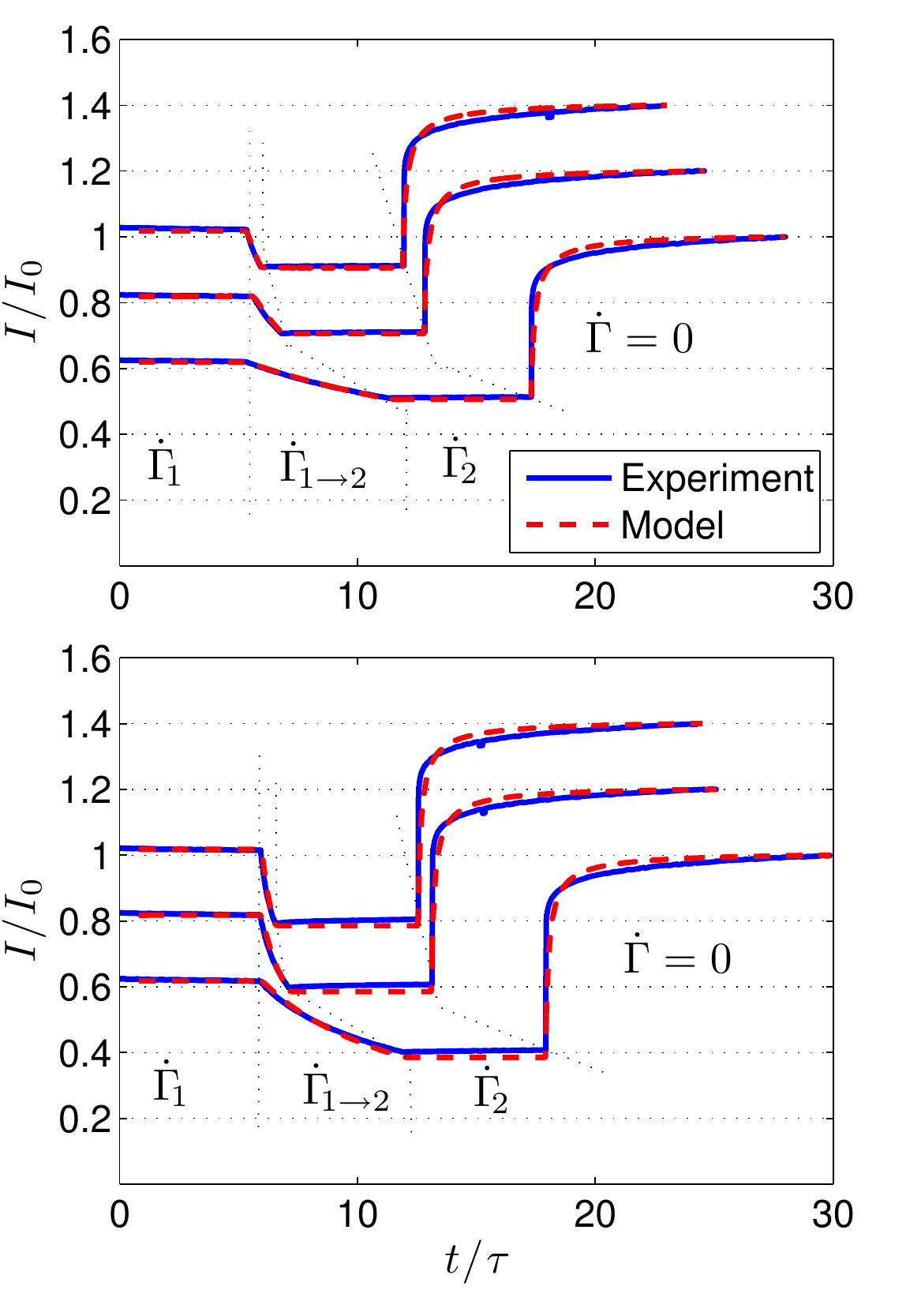}
  \caption{
    Transient ramp experiments from (top) $\tau \dot\Gamma_1=2500$ to $\tau \dot\Gamma_2=5000$,
    and (bottom) $\tau \dot\Gamma_1= 2500$ to $\tau \dot\Gamma_2=10000$ 
    with ramp times $t_R/\tau$ of 0.6, 1.2, and 6 (top to bottom).
    A vertical offset of 0.2 has been added between successive curves for clarity, 
    with the y-axis correct as displayed for the bottom curve.
    Current values are normalized by the static current measurement $I_0$. 
  }
  \label{fig:ramps}
\end{figure}

\textit{Results:}
The power law index $n$ was determined \textit{a priori} using the discrete simulation,
and all others were found using the fitting procedure described above.
The parameters used to generate all of the following plots, which satisfy all necessary
constraints, are
$Z_0 = 7$,
$Z_\infty = 2.41$,
$\tau = 50 \, s$,
$\beta = 0.009$,
$\alpha = -0.0089$,
$n = 1.34$, and
$k_1 = 0.0282\; S/m$.

The top of Fig \ref{fig:2:NLSSModelfit} shows the steady-state current predicted by
the model compared to experimental data at various shear rates.
The results indicate the model does an excellent job predicting the steady-state 
current measurements over the entire tested range of $\dot\Gamma$. 
The bottom of Fig \ref{fig:2:NLSSModelfit} demonstrates the behavior of the model 
regarding evolution of the coordination number from a common starting value for various 
scalar strain-rates, $\left|\mytensor{D}\right|$, in pure shearing (same data is shown 
in both plots, but under different axes definitions). 
Dotted lines show analytical solutions in the limiting cases of $\left|\mytensor{D}\right| \rightarrow 0$
and $\left|\mytensor{D}\right| \rightarrow \infty$ respectively.
The model's prediction that $Z$ exponentially decays to its steady state in the shear-dominated ($|\mytensor{D}| \gg (\beta\tau)^{-1}$)
limit agrees with previous simulations of attractive fluid-particle systems \cite{martys2009contact}.

The results in Fig \ref{fig:ramps} show the temporal evolution of the current under
imposed shearing normalized by the static current measurement.
The close agreement indicates that the model is capable of making accurate, quantitative predictions
for the transient behavior of the normalized current.
The model and the data match closely over the majority of the experiment,
and the model reproduces the final relaxation behavior correctly.

\textit{Discussion:}
We have demonstrated a model for the conductivity of sheared suspensions, by linking 
conductivity and flow to a common microstructural description.  
Although the model parameters were obtained using conductivity measurements, 
it is interesting to note the following points, which emphasize that the 
structure indeed plays the assumed role: 
(i) the best-fit parameters obtained from the calibration experiment obey 
constraints implied by a particle structure (\ref{eq:signConstraintC1}-\ref{eq:c1c3Constraint}), 
(ii) the parameters predict a reasonable static coordination number for a particle system, and 
(iii) the structural evolution model, on which the experimental agreement hinges, gives the 
same asymptotic behaviors as those observed in suspension simulations.
Conversely, our approach highlights the possibility of using conductivity measurements to 
infer discrete microstructural properties of systems, a notion also suggested in \cite{zhuang1995}.
It should be noted that a simpler constitutive
specialization of the $c_i$ functions using $n=0$ does not not capture the 
measured evolution of conductivity well.
A model using $n=0$ predicts exponential, rather than power-law, 
relaxation of fabric upon cessation of flow.
The resulting conductivity evolution cannot predict the observed behavior
shown in Fig \ref{fig:ramps}.
The new model (\ref{eq:KAtensor3D}, \ref{eq:EvolutionReduced}, \ref{eq:c1Function}-\ref{eq:c3Function})
can be applied as a quantitative tool for designing systems of flowing,
electrically-active pariculate suspensions, for example in a semi-solid flow battery architecture 
\cite{Duduta2011,Helal2014,Smith2014}.
Because the shear-induced conductivity loss can be strong even under moderate shear rates, 
the ability to predict this component of the performance envelope through a quantitative continuum 
model should enable subsequent geometric and flow protocol optimization.

\textit{Acknowledgements:}
The authors acknowledge support from the Joint Center for Energy Storage Research (JCESR), 
an Energy Innovation Hub funded by the U.S. Department of Energy, Office of Science, 
Basic Energy Science (BES).
The authors declare that there are no conflicts of interest.


\bibliography{main}

\begin{thebibliography}{43}%
\makeatletter
\providecommand \@ifxundefined [1]{%
 \@ifx{#1\undefined}
}%
\providecommand \@ifnum [1]{%
 \ifnum #1\expandafter \@firstoftwo
 \else \expandafter \@secondoftwo
 \fi
}%
\providecommand \@ifx [1]{%
 \ifx #1\expandafter \@firstoftwo
 \else \expandafter \@secondoftwo
 \fi
}%
\providecommand \natexlab [1]{#1}%
\providecommand \enquote  [1]{``#1''}%
\providecommand \bibnamefont  [1]{#1}%
\providecommand \bibfnamefont [1]{#1}%
\providecommand \citenamefont [1]{#1}%
\providecommand \href@noop [0]{\@secondoftwo}%
\providecommand \href [0]{\begingroup \@sanitize@url \@href}%
\providecommand \@href[1]{\@@startlink{#1}\@@href}%
\providecommand \@@href[1]{\endgroup#1\@@endlink}%
\providecommand \@sanitize@url [0]{\catcode `\\12\catcode `\$12\catcode
  `\&12\catcode `\#12\catcode `\^12\catcode `\_12\catcode `\%12\relax}%
\providecommand \@@startlink[1]{}%
\providecommand \@@endlink[0]{}%
\providecommand \url  [0]{\begingroup\@sanitize@url \@url }%
\providecommand \@url [1]{\endgroup\@href {#1}{\urlprefix }}%
\providecommand \urlprefix  [0]{URL }%
\providecommand \Eprint [0]{\href }%
\providecommand \doibase [0]{http://dx.doi.org/}%
\providecommand \selectlanguage [0]{\@gobble}%
\providecommand \bibinfo  [0]{\@secondoftwo}%
\providecommand \bibfield  [0]{\@secondoftwo}%
\providecommand \translation [1]{[#1]}%
\providecommand \BibitemOpen [0]{}%
\providecommand \bibitemStop [0]{}%
\providecommand \bibitemNoStop [0]{.\EOS\space}%
\providecommand \EOS [0]{\spacefactor3000\relax}%
\providecommand \BibitemShut  [1]{\csname bibitem#1\endcsname}%
\let\auto@bib@innerbib\@empty
\bibitem [{\citenamefont {Bischoff}\ \emph {et~al.}(2002)\citenamefont
  {Bischoff}, \citenamefont {Arruda},\ and\ \citenamefont
  {Grosh}}]{Bischoff2002}%
  \BibitemOpen
  \bibfield  {author} {\bibinfo {author} {\bibfnamefont {J.~E.}\ \bibnamefont
  {Bischoff}}, \bibinfo {author} {\bibfnamefont {E.~M.}\ \bibnamefont
  {Arruda}}, \ and\ \bibinfo {author} {\bibfnamefont {K.}~\bibnamefont
  {Grosh}},\ }\href {\doibase 10.1115/1.1485754} {\bibfield  {journal}
  {\bibinfo  {journal} {Journal of Applied Mechanics}\ }\textbf {\bibinfo
  {volume} {69}},\ \bibinfo {pages} {570} (\bibinfo {year} {2002})}\BibitemShut
  {NoStop}%
\bibitem [{\citenamefont {Deneweth}\ \emph {et~al.}(2013)\citenamefont
  {Deneweth}, \citenamefont {McLean},\ and\ \citenamefont
  {Arruda}}]{Deneweth2013}%
  \BibitemOpen
  \bibfield  {author} {\bibinfo {author} {\bibfnamefont {J.~M.}\ \bibnamefont
  {Deneweth}}, \bibinfo {author} {\bibfnamefont {S.~G.}\ \bibnamefont
  {McLean}}, \ and\ \bibinfo {author} {\bibfnamefont {E.~M.}\ \bibnamefont
  {Arruda}},\ }\href {\doibase 10.1016/j.jbiomech.2013.04.014} {\bibfield
  {journal} {\bibinfo  {journal} {Journal of biomechanics}\ }\textbf {\bibinfo
  {volume} {46}},\ \bibinfo {pages} {1604} (\bibinfo {year}
  {2013})}\BibitemShut {NoStop}%
\bibitem [{\citenamefont {Frederick}\ and\ \citenamefont
  {Armstrong}(2007)}]{Frederick2007}%
  \BibitemOpen
  \bibfield  {author} {\bibinfo {author} {\bibfnamefont {C.~O.}\ \bibnamefont
  {Frederick}}\ and\ \bibinfo {author} {\bibfnamefont {P.}~\bibnamefont
  {Armstrong}},\ }\href@noop {} {\bibfield  {journal} {\bibinfo  {journal}
  {Materials at High Temperatures}\ }\textbf {\bibinfo {volume} {24}},\
  \bibinfo {pages} {1} (\bibinfo {year} {2007})}\BibitemShut {NoStop}%
\bibitem [{\citenamefont {Az\'{e}ma}\ \emph {et~al.}(2012)\citenamefont
  {Az\'{e}ma}, \citenamefont {Estrada},\ and\ \citenamefont
  {Radjai}}]{Azema2012}%
  \BibitemOpen
  \bibfield  {author} {\bibinfo {author} {\bibfnamefont {E.}~\bibnamefont
  {Az\'{e}ma}}, \bibinfo {author} {\bibfnamefont {N.}~\bibnamefont {Estrada}},
  \ and\ \bibinfo {author} {\bibfnamefont {F.}~\bibnamefont {Radjai}},\ }\href
  {http://pre.aps.org/abstract/PRE/v86/i4/e041301} {\bibfield  {journal}
  {\bibinfo  {journal} {Physical Review E}\ }\textbf {\bibinfo {volume} {86}},\
  \bibinfo {pages} {041301} (\bibinfo {year} {2012})}\BibitemShut {NoStop}%
\bibitem [{\citenamefont {da~Cruz}\ \emph {et~al.}(2005)\citenamefont
  {da~Cruz}, \citenamefont {Emam}, \citenamefont {Prochnow}, \citenamefont
  {Roux},\ and\ \citenamefont {Chevoir}}]{DaCruz2005}%
  \BibitemOpen
  \bibfield  {author} {\bibinfo {author} {\bibfnamefont {F.}~\bibnamefont
  {da~Cruz}}, \bibinfo {author} {\bibfnamefont {S.}~\bibnamefont {Emam}},
  \bibinfo {author} {\bibfnamefont {M.}~\bibnamefont {Prochnow}}, \bibinfo
  {author} {\bibfnamefont {J.-N.}\ \bibnamefont {Roux}}, \ and\ \bibinfo
  {author} {\bibfnamefont {F.}~\bibnamefont {Chevoir}},\ }\href {\doibase
  10.1103/PhysRevE.72.021309} {\bibfield  {journal} {\bibinfo  {journal}
  {Physical Review E}\ }\textbf {\bibinfo {volume} {72}},\ \bibinfo {pages}
  {021309} (\bibinfo {year} {2005})}\BibitemShut {NoStop}%
\bibitem [{\citenamefont {Mehrabadi}\ \emph {et~al.}(1982)\citenamefont
  {Mehrabadi}, \citenamefont {Nemat-Nasser},\ and\ \citenamefont
  {Oda}}]{Mehrabadi1982}%
  \BibitemOpen
  \bibfield  {author} {\bibinfo {author} {\bibfnamefont {M.~M.}\ \bibnamefont
  {Mehrabadi}}, \bibinfo {author} {\bibfnamefont {S.}~\bibnamefont
  {Nemat-Nasser}}, \ and\ \bibinfo {author} {\bibfnamefont {M.}~\bibnamefont
  {Oda}},\ }\href@noop {} {\bibfield  {journal} {\bibinfo  {journal}
  {International Journal for Numerical and Analytical Methods in Geomechanics}\
  }\textbf {\bibinfo {volume} {6}},\ \bibinfo {pages} {95} (\bibinfo {year}
  {1982})}\BibitemShut {NoStop}%
\bibitem [{\citenamefont {Oda}\ \emph {et~al.}(1982)\citenamefont {Oda},
  \citenamefont {Nemat-Nasser},\ and\ \citenamefont {Mehrabadi}}]{Oda1982}%
  \BibitemOpen
  \bibfield  {author} {\bibinfo {author} {\bibfnamefont {M.}~\bibnamefont
  {Oda}}, \bibinfo {author} {\bibfnamefont {S.}~\bibnamefont {Nemat-Nasser}}, \
  and\ \bibinfo {author} {\bibfnamefont {M.~M.}\ \bibnamefont {Mehrabadi}},\
  }\href {http://onlinelibrary.wiley.com/doi/10.1002/nag.1610060106/abstract}
  {\bibfield  {journal} {\bibinfo  {journal} {International Journal for
  Numerical and Analytical Methods in Geomechanics}\ }\textbf {\bibinfo
  {volume} {6}},\ \bibinfo {pages} {77} (\bibinfo {year} {1982})}\BibitemShut
  {NoStop}%
\bibitem [{\citenamefont {Radjai}\ \emph {et~al.}(2012)\citenamefont {Radjai},
  \citenamefont {Delenne}, \citenamefont {Az\'{e}ma},\ and\ \citenamefont
  {Roux}}]{Radjai2012}%
  \BibitemOpen
  \bibfield  {author} {\bibinfo {author} {\bibfnamefont {F.}~\bibnamefont
  {Radjai}}, \bibinfo {author} {\bibfnamefont {J.-Y.}\ \bibnamefont {Delenne}},
  \bibinfo {author} {\bibfnamefont {E.}~\bibnamefont {Az\'{e}ma}}, \ and\
  \bibinfo {author} {\bibfnamefont {S.}~\bibnamefont {Roux}},\ }\href {\doibase
  10.1007/s10035-012-0321-8} {\bibfield  {journal} {\bibinfo  {journal}
  {Granular Matter}\ }\textbf {\bibinfo {volume} {14}},\ \bibinfo {pages} {259}
  (\bibinfo {year} {2012})}\BibitemShut {NoStop}%
\bibitem [{\citenamefont {Satake}(1978)}]{Satake1978}%
  \BibitemOpen
  \bibfield  {author} {\bibinfo {author} {\bibfnamefont {M.}~\bibnamefont
  {Satake}},\ }\href@noop {} {\bibfield  {journal} {\bibinfo  {journal}
  {Continuum Mechanical and Statistical Approaches in the Mechanics of Granular
  Materials}\ ,\ \bibinfo {pages} {47}} (\bibinfo {year} {1978})}\BibitemShut
  {NoStop}%
\bibitem [{\citenamefont {Sun}\ and\ \citenamefont
  {Sundaresan}(2011)}]{Sun2011}%
  \BibitemOpen
  \bibfield  {author} {\bibinfo {author} {\bibfnamefont {J.}~\bibnamefont
  {Sun}}\ and\ \bibinfo {author} {\bibfnamefont {S.}~\bibnamefont
  {Sundaresan}},\ }\href {\doibase 10.1017/jfm.2011.251} {\bibfield  {journal}
  {\bibinfo  {journal} {Journal of Fluid Mechanics}\ }\textbf {\bibinfo
  {volume} {682}},\ \bibinfo {pages} {590} (\bibinfo {year}
  {2011})}\BibitemShut {NoStop}%
\bibitem [{\citenamefont {Stephen}\ and\ \citenamefont
  {Straley}(1974)}]{Stephen1974}%
  \BibitemOpen
  \bibfield  {author} {\bibinfo {author} {\bibfnamefont {M.~J.}\ \bibnamefont
  {Stephen}}\ and\ \bibinfo {author} {\bibfnamefont {J.~P.}\ \bibnamefont
  {Straley}},\ }\href@noop {} {\bibfield  {journal} {\bibinfo  {journal}
  {Reviews of Modern Physics}\ }\textbf {\bibinfo {volume} {46}},\ \bibinfo
  {pages} {617} (\bibinfo {year} {1974})}\BibitemShut {NoStop}%
\bibitem [{\citenamefont {Grenard}\ \emph {et~al.}(2011)\citenamefont
  {Grenard}, \citenamefont {Taberlet},\ and\ \citenamefont
  {Manneville}}]{Grenard2011}%
  \BibitemOpen
  \bibfield  {author} {\bibinfo {author} {\bibfnamefont {V.}~\bibnamefont
  {Grenard}}, \bibinfo {author} {\bibfnamefont {N.}~\bibnamefont {Taberlet}}, \
  and\ \bibinfo {author} {\bibfnamefont {S.}~\bibnamefont {Manneville}},\
  }\href {\doibase 10.1039/C0SM01515F} {\bibfield  {journal} {\bibinfo
  {journal} {Soft Matter}\ }\textbf {\bibinfo {volume} {7}},\ \bibinfo {pages}
  {3920} (\bibinfo {year} {2011})}\BibitemShut {NoStop}%
\bibitem [{\citenamefont {Grenard}\ \emph {et~al.}(2014)\citenamefont
  {Grenard}, \citenamefont {Divoux}, \citenamefont {Taberlet},\ and\
  \citenamefont {Manneville}}]{Grenard2014}%
  \BibitemOpen
  \bibfield  {author} {\bibinfo {author} {\bibfnamefont {V.}~\bibnamefont
  {Grenard}}, \bibinfo {author} {\bibfnamefont {T.}~\bibnamefont {Divoux}},
  \bibinfo {author} {\bibfnamefont {N.}~\bibnamefont {Taberlet}}, \ and\
  \bibinfo {author} {\bibfnamefont {S.}~\bibnamefont {Manneville}},\ }\href
  {\doibase 10.1039/C3SM52548A} {\bibfield  {journal} {\bibinfo  {journal}
  {Soft Matter}\ }\textbf {\bibinfo {volume} {10}},\ \bibinfo {pages} {1555}
  (\bibinfo {year} {2014})}\BibitemShut {NoStop}%
\bibitem [{\citenamefont {Koumakis}\ \emph {et~al.}(2012)\citenamefont
  {Koumakis}, \citenamefont {Laurati}, \citenamefont {Egelhaaf}, \citenamefont
  {Brady},\ and\ \citenamefont {Petekidis}}]{Koumakis2012}%
  \BibitemOpen
  \bibfield  {author} {\bibinfo {author} {\bibfnamefont {N.}~\bibnamefont
  {Koumakis}}, \bibinfo {author} {\bibfnamefont {M.}~\bibnamefont {Laurati}},
  \bibinfo {author} {\bibfnamefont {S.~U.}\ \bibnamefont {Egelhaaf}}, \bibinfo
  {author} {\bibfnamefont {J.~F.}\ \bibnamefont {Brady}}, \ and\ \bibinfo
  {author} {\bibfnamefont {G.}~\bibnamefont {Petekidis}},\ }\href {\doibase
  10.1103/PhysRevLett.108.098303} {\bibfield  {journal} {\bibinfo  {journal}
  {Physical Review Letters}\ }\textbf {\bibinfo {volume} {108}},\ \bibinfo
  {pages} {098303} (\bibinfo {year} {2012})}\BibitemShut {NoStop}%
\bibitem [{\citenamefont {Mohraz}\ and\ \citenamefont
  {Solomon}(2005)}]{Mohraz2005}%
  \BibitemOpen
  \bibfield  {author} {\bibinfo {author} {\bibfnamefont {A.}~\bibnamefont
  {Mohraz}}\ and\ \bibinfo {author} {\bibfnamefont {M.~J.}\ \bibnamefont
  {Solomon}},\ }\href {\doibase 10.1122/1.1895799} {\bibfield  {journal}
  {\bibinfo  {journal} {Journal of Rheology}\ }\textbf {\bibinfo {volume}
  {49}},\ \bibinfo {pages} {657} (\bibinfo {year} {2005})}\BibitemShut
  {NoStop}%
\bibitem [{\citenamefont {Negi}\ and\ \citenamefont {Osuji}(2009)}]{Negi2009}%
  \BibitemOpen
  \bibfield  {author} {\bibinfo {author} {\bibfnamefont {A.~S.}\ \bibnamefont
  {Negi}}\ and\ \bibinfo {author} {\bibfnamefont {C.~O.}\ \bibnamefont
  {Osuji}},\ }\href {\doibase 10.1007/s00397-008-0341-9} {\bibfield  {journal}
  {\bibinfo  {journal} {Rheologica Acta}\ }\textbf {\bibinfo {volume} {48}},\
  \bibinfo {pages} {871} (\bibinfo {year} {2009})}\BibitemShut {NoStop}%
\bibitem [{\citenamefont {Osuji}\ \emph {et~al.}(2008)\citenamefont {Osuji},
  \citenamefont {Kim},\ and\ \citenamefont {Weitz}}]{Osuji2008}%
  \BibitemOpen
  \bibfield  {author} {\bibinfo {author} {\bibfnamefont {C.~O.}\ \bibnamefont
  {Osuji}}, \bibinfo {author} {\bibfnamefont {C.}~\bibnamefont {Kim}}, \ and\
  \bibinfo {author} {\bibfnamefont {D.~A.}\ \bibnamefont {Weitz}},\ }\href
  {\doibase 10.1103/PhysRevE.77.060402} {\bibfield  {journal} {\bibinfo
  {journal} {Physical Review E - Statistical, Nonlinear, and Soft Matter
  Physics}\ }\textbf {\bibinfo {volume} {77}},\ \bibinfo {pages} {8} (\bibinfo
  {year} {2008})}\BibitemShut {NoStop}%
\bibitem [{\citenamefont {Osuji}\ and\ \citenamefont
  {Weitz}(2008)}]{Osuji2008a}%
  \BibitemOpen
  \bibfield  {author} {\bibinfo {author} {\bibfnamefont {C.~O.}\ \bibnamefont
  {Osuji}}\ and\ \bibinfo {author} {\bibfnamefont {D.~A.}\ \bibnamefont
  {Weitz}},\ }\href {\doibase 10.1039/B716324J} {\bibfield  {journal} {\bibinfo
   {journal} {Soft Matter}\ }\textbf {\bibinfo {volume} {4}},\ \bibinfo {pages}
  {1388} (\bibinfo {year} {2008})}\BibitemShut {NoStop}%
\bibitem [{\citenamefont {Park}\ and\ \citenamefont {Ahn}(2013)}]{Park2013}%
  \BibitemOpen
  \bibfield  {author} {\bibinfo {author} {\bibfnamefont {J.~D.}\ \bibnamefont
  {Park}}\ and\ \bibinfo {author} {\bibfnamefont {K.~H.}\ \bibnamefont {Ahn}},\
  }\href {\doibase 10.1039/C3SM52090K} {\bibfield  {journal} {\bibinfo
  {journal} {Soft Matter}\ }\textbf {\bibinfo {volume} {9}},\ \bibinfo {pages}
  {11650} (\bibinfo {year} {2013})}\BibitemShut {NoStop}%
\bibitem [{\citenamefont {Santos}\ \emph {et~al.}(2013)\citenamefont {Santos},
  \citenamefont {Campanella},\ and\ \citenamefont {Carignano}}]{Santos2013}%
  \BibitemOpen
  \bibfield  {author} {\bibinfo {author} {\bibfnamefont {P.~H.~S.}\
  \bibnamefont {Santos}}, \bibinfo {author} {\bibfnamefont {O.~H.}\
  \bibnamefont {Campanella}}, \ and\ \bibinfo {author} {\bibfnamefont {M.~A.}\
  \bibnamefont {Carignano}},\ }\href {\doibase 10.1039/c2sm26585k} {\bibfield
  {journal} {\bibinfo  {journal} {Soft Matter}\ }\textbf {\bibinfo {volume}
  {9}},\ \bibinfo {pages} {709} (\bibinfo {year} {2013})},\ \Eprint
  {http://arxiv.org/abs/1307.3615} {arXiv:1307.3615} \BibitemShut {NoStop}%
\bibitem [{\citenamefont {Trappe}\ \emph {et~al.}(2001)\citenamefont {Trappe},
  \citenamefont {Prasad}, \citenamefont {Cipelletti}, \citenamefont {Segre},\
  and\ \citenamefont {Weitz}}]{Trappe2001}%
  \BibitemOpen
  \bibfield  {author} {\bibinfo {author} {\bibfnamefont {V.}~\bibnamefont
  {Trappe}}, \bibinfo {author} {\bibfnamefont {V.}~\bibnamefont {Prasad}},
  \bibinfo {author} {\bibfnamefont {L.}~\bibnamefont {Cipelletti}}, \bibinfo
  {author} {\bibfnamefont {P.~N.}\ \bibnamefont {Segre}}, \ and\ \bibinfo
  {author} {\bibfnamefont {D.~A.}\ \bibnamefont {Weitz}},\ }\href {\doibase
  10.1038/35081021} {\bibfield  {journal} {\bibinfo  {journal} {Nature}\
  }\textbf {\bibinfo {volume} {411}},\ \bibinfo {pages} {772} (\bibinfo {year}
  {2001})}\BibitemShut {NoStop}%
\bibitem [{\citenamefont {Trappe}\ and\ \citenamefont
  {Weitz}(2000)}]{Trappe2000}%
  \BibitemOpen
  \bibfield  {author} {\bibinfo {author} {\bibfnamefont {V.}~\bibnamefont
  {Trappe}}\ and\ \bibinfo {author} {\bibfnamefont {D.~A.}\ \bibnamefont
  {Weitz}},\ }\href {\doibase 10.1103/PhysRevLett.85.449} {\bibfield  {journal}
  {\bibinfo  {journal} {Physical Review Letters}\ }\textbf {\bibinfo {volume}
  {85}},\ \bibinfo {pages} {449} (\bibinfo {year} {2000})}\BibitemShut
  {NoStop}%
\bibitem [{\citenamefont {Duduta}\ \emph {et~al.}(2011)\citenamefont {Duduta},
  \citenamefont {Ho}, \citenamefont {Wood}, \citenamefont {Limthongkul},
  \citenamefont {Brunini}, \citenamefont {Carter},\ and\ \citenamefont
  {Chiang}}]{Duduta2011}%
  \BibitemOpen
  \bibfield  {author} {\bibinfo {author} {\bibfnamefont {M.}~\bibnamefont
  {Duduta}}, \bibinfo {author} {\bibfnamefont {B.}~\bibnamefont {Ho}}, \bibinfo
  {author} {\bibfnamefont {V.~C.}\ \bibnamefont {Wood}}, \bibinfo {author}
  {\bibfnamefont {P.}~\bibnamefont {Limthongkul}}, \bibinfo {author}
  {\bibfnamefont {V.~E.}\ \bibnamefont {Brunini}}, \bibinfo {author}
  {\bibfnamefont {W.~C.}\ \bibnamefont {Carter}}, \ and\ \bibinfo {author}
  {\bibfnamefont {Y.-M.}\ \bibnamefont {Chiang}},\ }\href {\doibase
  10.1002/aenm.201100152} {\bibfield  {journal} {\bibinfo  {journal} {Advanced
  Energy Materials}\ }\textbf {\bibinfo {volume} {1}},\ \bibinfo {pages} {511}
  (\bibinfo {year} {2011})}\BibitemShut {NoStop}%
\bibitem [{\citenamefont {Youssry}\ \emph {et~al.}(2013)\citenamefont
  {Youssry}, \citenamefont {Madec}, \citenamefont {Soudan}, \citenamefont
  {Cerbelaud}, \citenamefont {Guyomard},\ and\ \citenamefont
  {Lestriez}}]{Youssry2013}%
  \BibitemOpen
  \bibfield  {author} {\bibinfo {author} {\bibfnamefont {M.}~\bibnamefont
  {Youssry}}, \bibinfo {author} {\bibfnamefont {L.}~\bibnamefont {Madec}},
  \bibinfo {author} {\bibfnamefont {P.}~\bibnamefont {Soudan}}, \bibinfo
  {author} {\bibfnamefont {M.}~\bibnamefont {Cerbelaud}}, \bibinfo {author}
  {\bibfnamefont {D.}~\bibnamefont {Guyomard}}, \ and\ \bibinfo {author}
  {\bibfnamefont {B.}~\bibnamefont {Lestriez}},\ }\href {\doibase
  10.1039/c3cp51371h} {\bibfield  {journal} {\bibinfo  {journal} {Physical
  Chemistry Chemical Physics}\ }\textbf {\bibinfo {volume} {15}},\ \bibinfo
  {pages} {14476} (\bibinfo {year} {2013})}\BibitemShut {NoStop}%
\bibitem [{\citenamefont {Alig}\ \emph {et~al.}(2007)\citenamefont {Alig},
  \citenamefont {Skipa}, \citenamefont {Engel}, \citenamefont {Lellinger},
  \citenamefont {Pegel},\ and\ \citenamefont {P\"{o}tschke}}]{Alig2007}%
  \BibitemOpen
  \bibfield  {author} {\bibinfo {author} {\bibfnamefont {I.}~\bibnamefont
  {Alig}}, \bibinfo {author} {\bibfnamefont {T.}~\bibnamefont {Skipa}},
  \bibinfo {author} {\bibfnamefont {M.}~\bibnamefont {Engel}}, \bibinfo
  {author} {\bibfnamefont {D.}~\bibnamefont {Lellinger}}, \bibinfo {author}
  {\bibfnamefont {S.}~\bibnamefont {Pegel}}, \ and\ \bibinfo {author}
  {\bibfnamefont {P.}~\bibnamefont {P\"{o}tschke}},\ }\href {\doibase
  10.1002/pssb.200776138} {\bibfield  {journal} {\bibinfo  {journal} {Physica
  Status Solidi (B)}\ }\textbf {\bibinfo {volume} {244}},\ \bibinfo {pages}
  {4223} (\bibinfo {year} {2007})}\BibitemShut {NoStop}%
\bibitem [{\citenamefont {Amari}\ and\ \citenamefont
  {Watanabe}(1990)}]{Amari1990}%
  \BibitemOpen
  \bibfield  {author} {\bibinfo {author} {\bibfnamefont {T.}~\bibnamefont
  {Amari}}\ and\ \bibinfo {author} {\bibfnamefont {K.}~\bibnamefont
  {Watanabe}},\ }\href {\doibase 10.1122/1.550124} {\bibfield  {journal}
  {\bibinfo  {journal} {Journal of Rheology}\ }\textbf {\bibinfo {volume}
  {34}},\ \bibinfo {pages} {207} (\bibinfo {year} {1990})}\BibitemShut
  {NoStop}%
\bibitem [{\citenamefont {Bauhofer}\ \emph {et~al.}(2010)\citenamefont
  {Bauhofer}, \citenamefont {Schulz}, \citenamefont {Eken}, \citenamefont
  {Skipa}, \citenamefont {Lellinger}, \citenamefont {Alig}, \citenamefont
  {Tozzi},\ and\ \citenamefont {Klingenberg}}]{Bauhofer2010}%
  \BibitemOpen
  \bibfield  {author} {\bibinfo {author} {\bibfnamefont {W.}~\bibnamefont
  {Bauhofer}}, \bibinfo {author} {\bibfnamefont {S.}~\bibnamefont {Schulz}},
  \bibinfo {author} {\bibfnamefont {A.~E.}\ \bibnamefont {Eken}}, \bibinfo
  {author} {\bibfnamefont {T.}~\bibnamefont {Skipa}}, \bibinfo {author}
  {\bibfnamefont {D.}~\bibnamefont {Lellinger}}, \bibinfo {author}
  {\bibfnamefont {I.}~\bibnamefont {Alig}}, \bibinfo {author} {\bibfnamefont
  {E.}~\bibnamefont {Tozzi}}, \ and\ \bibinfo {author} {\bibfnamefont
  {D.}~\bibnamefont {Klingenberg}},\ }\href {\doibase
  10.1016/j.polymer.2010.09.013} {\bibfield  {journal} {\bibinfo  {journal}
  {Polymer}\ }\textbf {\bibinfo {volume} {51}},\ \bibinfo {pages} {5024}
  (\bibinfo {year} {2010})}\BibitemShut {NoStop}%
\bibitem [{\citenamefont {Schulz}\ and\ \citenamefont
  {Bauhofer}(2010)}]{Schulz2010}%
  \BibitemOpen
  \bibfield  {author} {\bibinfo {author} {\bibfnamefont {S.}~\bibnamefont
  {Schulz}}\ and\ \bibinfo {author} {\bibfnamefont {W.}~\bibnamefont
  {Bauhofer}},\ }\href {\doibase 10.1016/j.polymer.2010.09.027} {\bibfield
  {journal} {\bibinfo  {journal} {Polymer}\ }\textbf {\bibinfo {volume} {51}},\
  \bibinfo {pages} {5500} (\bibinfo {year} {2010})}\BibitemShut {NoStop}%
\bibitem [{\citenamefont {Smith}\ \emph {et~al.}(2014)\citenamefont {Smith},
  \citenamefont {Chiang},\ and\ \citenamefont {{Craig Carter}}}]{Smith2014}%
  \BibitemOpen
  \bibfield  {author} {\bibinfo {author} {\bibfnamefont {K.~C.}\ \bibnamefont
  {Smith}}, \bibinfo {author} {\bibfnamefont {Y.-M.}\ \bibnamefont {Chiang}}, \
  and\ \bibinfo {author} {\bibfnamefont {W.}~\bibnamefont {{Craig Carter}}},\
  }\href {\doibase 10.1149/2.011404jes} {\bibfield  {journal} {\bibinfo
  {journal} {Journal of the Electrochemical Society}\ }\textbf {\bibinfo
  {volume} {161}},\ \bibinfo {pages} {A486} (\bibinfo {year}
  {2014})}\BibitemShut {NoStop}%
\bibitem [{\citenamefont {Hoekstra}\ \emph {et~al.}(2003)\citenamefont
  {Hoekstra}, \citenamefont {Vermant}, \citenamefont {Mewis},\ and\
  \citenamefont {Fuller}}]{Hoekstra2003}%
  \BibitemOpen
  \bibfield  {author} {\bibinfo {author} {\bibfnamefont {H.}~\bibnamefont
  {Hoekstra}}, \bibinfo {author} {\bibfnamefont {J.}~\bibnamefont {Vermant}},
  \bibinfo {author} {\bibfnamefont {J.}~\bibnamefont {Mewis}}, \ and\ \bibinfo
  {author} {\bibfnamefont {G.~G.}\ \bibnamefont {Fuller}},\ }\href
  {http://pubs.acs.org/doi/abs/10.1021/la034582k} {\bibfield  {journal}
  {\bibinfo  {journal} {Langmuir}\ ,\ \bibinfo {pages} {9134}} (\bibinfo {year}
  {2003})}\BibitemShut {NoStop}%
\bibitem [{\citenamefont {Morris}\ and\ \citenamefont
  {Katyal}(2002)}]{Morris2002}%
  \BibitemOpen
  \bibfield  {author} {\bibinfo {author} {\bibfnamefont {J.~F.}\ \bibnamefont
  {Morris}}\ and\ \bibinfo {author} {\bibfnamefont {B.}~\bibnamefont
  {Katyal}},\ }\href {\doibase 10.1063/1.1476745} {\bibfield  {journal}
  {\bibinfo  {journal} {Physics of Fluids}\ }\textbf {\bibinfo {volume} {14}},\
  \bibinfo {pages} {1920} (\bibinfo {year} {2002})}\BibitemShut {NoStop}%
\bibitem [{\citenamefont {Vermant}\ and\ \citenamefont
  {Solomon}(2005)}]{Vermant2005}%
  \BibitemOpen
  \bibfield  {author} {\bibinfo {author} {\bibfnamefont {J.}~\bibnamefont
  {Vermant}}\ and\ \bibinfo {author} {\bibfnamefont {M.~J.}\ \bibnamefont
  {Solomon}},\ }\href {\doibase 10.1088/0953-8984/17/4/R02} {\bibfield
  {journal} {\bibinfo  {journal} {Journal of Physics: Condensed Matter}\
  }\textbf {\bibinfo {volume} {17}},\ \bibinfo {pages} {R187} (\bibinfo {year}
  {2005})}\BibitemShut {NoStop}%
\bibitem [{\citenamefont {Olsen}\ and\ \citenamefont
  {Kamrin}(2015)}]{Olsen2015}%
  \BibitemOpen
  \bibfield  {author} {\bibinfo {author} {\bibfnamefont {T.}~\bibnamefont
  {Olsen}}\ and\ \bibinfo {author} {\bibfnamefont {K.}~\bibnamefont {Kamrin}},\
  }\href@noop {} {\bibfield  {journal} {\bibinfo  {journal} {Soft Matter}\
  }\textbf {\bibinfo {volume} {11}},\ \bibinfo {pages} {3875} (\bibinfo {year}
  {2015})}\BibitemShut {NoStop}%
\bibitem [{\citenamefont {Hand}(1962)}]{Hand1961a}%
  \BibitemOpen
  \bibfield  {author} {\bibinfo {author} {\bibfnamefont {G.~L.}\ \bibnamefont
  {Hand}},\ }\href {\doibase 10.1017/S0022112062000476} {\bibfield  {journal}
  {\bibinfo  {journal} {Journal of Fluid Mechanics}\ }\textbf {\bibinfo
  {volume} {13}},\ \bibinfo {pages} {33} (\bibinfo {year} {1962})}\BibitemShut
  {NoStop}%
\bibitem [{\citenamefont {Rivlin}(1955)}]{Rivlin1955}%
  \BibitemOpen
  \bibfield  {author} {\bibinfo {author} {\bibfnamefont {R.~S.}\ \bibnamefont
  {Rivlin}},\ }\href@noop {} {\bibfield  {journal} {\bibinfo  {journal}
  {Journal of Rational Mechanics and Analysis}\ }\textbf {\bibinfo {volume}
  {4}},\ \bibinfo {pages} {681} (\bibinfo {year} {1955})}\BibitemShut {NoStop}%
\bibitem [{\citenamefont {Marsden}\ and\ \citenamefont
  {Hughes}(1994)}]{Marsden1994}%
  \BibitemOpen
  \bibfield  {author} {\bibinfo {author} {\bibfnamefont {J.~E.}\ \bibnamefont
  {Marsden}}\ and\ \bibinfo {author} {\bibfnamefont {T.~J.}\ \bibnamefont
  {Hughes}},\ }\href@noop {} {\emph {\bibinfo {title} {Mathematical foundations
  of elasticity}}}\ (\bibinfo  {publisher} {Courier Corporation},\ \bibinfo
  {year} {1994})\BibitemShut {NoStop}%
\bibitem [{\citenamefont {Gurtin}\ \emph {et~al.}(2010)\citenamefont {Gurtin},
  \citenamefont {Fried},\ and\ \citenamefont {Anand}}]{Gurtin2010}%
  \BibitemOpen
  \bibfield  {author} {\bibinfo {author} {\bibfnamefont {M.~E.}\ \bibnamefont
  {Gurtin}}, \bibinfo {author} {\bibfnamefont {E.}~\bibnamefont {Fried}}, \
  and\ \bibinfo {author} {\bibfnamefont {L.}~\bibnamefont {Anand}},\
  }\href@noop {} {\emph {\bibinfo {title} {The Mechanics and Thermodynamics of
  Continua}}}\ (\bibinfo  {publisher} {Cambridge University Press},\ \bibinfo
  {year} {2010})\BibitemShut {NoStop}%
\bibitem [{\citenamefont {Jr}\ and\ \citenamefont {Sander}(1981)}]{Jr1981}%
  \BibitemOpen
  \bibfield  {author} {\bibinfo {author} {\bibfnamefont {T.~W.}\ \bibnamefont
  {Jr}}\ and\ \bibinfo {author} {\bibfnamefont {L.}~\bibnamefont {Sander}},\
  }\href {http://journals.aps.org/prl/abstract/10.1103/PhysRevLett.47.1400}
  {\bibfield  {journal} {\bibinfo  {journal} {Physical Review Letters}\
  }\textbf {\bibinfo {volume} {47}} (\bibinfo {year} {1981})}\BibitemShut
  {NoStop}%
\bibitem [{\citenamefont {Happel}\ and\ \citenamefont
  {Brenner}(2012)}]{Happel2012}%
  \BibitemOpen
  \bibfield  {author} {\bibinfo {author} {\bibfnamefont {J.}~\bibnamefont
  {Happel}}\ and\ \bibinfo {author} {\bibfnamefont {H.}~\bibnamefont
  {Brenner}},\ }\href@noop {} {\emph {\bibinfo {title} {Low Reynolds number
  hydrodynamics: with special applications to particulate media}}},\
  Vol.~\bibinfo {volume} {1}\ (\bibinfo  {publisher} {Springer Science \&
  Business Media},\ \bibinfo {year} {2012})\BibitemShut {NoStop}%
\bibitem [{\citenamefont {Helal}\ \emph {et~al.}(tion)\citenamefont {Helal},
  \citenamefont {Divoux}, \citenamefont {Chen}, \citenamefont {Chiang},\ and\
  \citenamefont {McKinley}}]{Helal201x}%
  \BibitemOpen
  \bibfield  {author} {\bibinfo {author} {\bibfnamefont {A.}~\bibnamefont
  {Helal}}, \bibinfo {author} {\bibfnamefont {T.}~\bibnamefont {Divoux}},
  \bibinfo {author} {\bibfnamefont {X.}~\bibnamefont {Chen}}, \bibinfo {author}
  {\bibfnamefont {Y.-M.}\ \bibnamefont {Chiang}}, \ and\ \bibinfo {author}
  {\bibfnamefont {G.~H.}\ \bibnamefont {McKinley}},\ }\href@noop {} {\bibfield
  {journal} {\bibinfo  {journal} {Phys. Rev. Applied}\ } (\bibinfo {year}
  {InPreparation})}\BibitemShut {NoStop}%
\bibitem [{\citenamefont {Martys}\ \emph {et~al.}(2009)\citenamefont {Martys},
  \citenamefont {Lootens}, \citenamefont {George},\ and\ \citenamefont
  {H{\'e}braud}}]{martys2009contact}%
  \BibitemOpen
  \bibfield  {author} {\bibinfo {author} {\bibfnamefont {N.~S.}\ \bibnamefont
  {Martys}}, \bibinfo {author} {\bibfnamefont {D.}~\bibnamefont {Lootens}},
  \bibinfo {author} {\bibfnamefont {W.}~\bibnamefont {George}}, \ and\ \bibinfo
  {author} {\bibfnamefont {P.}~\bibnamefont {H{\'e}braud}},\ }\href@noop {}
  {\bibfield  {journal} {\bibinfo  {journal} {Physical Review E}\ }\textbf
  {\bibinfo {volume} {80}},\ \bibinfo {pages} {031401} (\bibinfo {year}
  {2009})}\BibitemShut {NoStop}%
\bibitem [{\citenamefont {Zhuang}\ \emph {et~al.}(1995)\citenamefont {Zhuang},
  \citenamefont {Didwania},\ and\ \citenamefont {Goddard}}]{zhuang1995}%
  \BibitemOpen
  \bibfield  {author} {\bibinfo {author} {\bibfnamefont {X.}~\bibnamefont
  {Zhuang}}, \bibinfo {author} {\bibfnamefont {A.}~\bibnamefont {Didwania}}, \
  and\ \bibinfo {author} {\bibfnamefont {J.}~\bibnamefont {Goddard}},\
  }\href@noop {} {\bibfield  {journal} {\bibinfo  {journal} {Journal of
  Computational physics}\ }\textbf {\bibinfo {volume} {121}},\ \bibinfo {pages}
  {331} (\bibinfo {year} {1995})}\BibitemShut {NoStop}%
\bibitem [{\citenamefont {Helal}\ \emph {et~al.}(2014)\citenamefont {Helal},
  \citenamefont {Smith}, \citenamefont {Fan}, \citenamefont {Chen},
  \citenamefont {Nobrega}, \citenamefont {Chiang},\ and\ \citenamefont
  {McKinley}}]{Helal2014}%
  \BibitemOpen
  \bibfield  {author} {\bibinfo {author} {\bibfnamefont {A.}~\bibnamefont
  {Helal}}, \bibinfo {author} {\bibfnamefont {K.}~\bibnamefont {Smith}},
  \bibinfo {author} {\bibfnamefont {F.}~\bibnamefont {Fan}}, \bibinfo {author}
  {\bibfnamefont {X.~W.}\ \bibnamefont {Chen}}, \bibinfo {author}
  {\bibfnamefont {J.~M.}\ \bibnamefont {Nobrega}}, \bibinfo {author}
  {\bibfnamefont {Y.-M.}\ \bibnamefont {Chiang}}, \ and\ \bibinfo {author}
  {\bibfnamefont {G.~H.}\ \bibnamefont {McKinley}},\ }\href
  {http://www.rheology.org/sor14a/ViewPaper.aspx?ID=282} {\enquote {\bibinfo
  {title} {Study of the rheology and wall slip of carbon black suspensions for
  semi-solid flow batteries},}\ }\bibinfo {howpublished} {The Society of
  Rheology 86th Annual Meeting} (\bibinfo {year} {2014})\BibitemShut {NoStop}%
\end{thebibliography}%
\bibliographystyle{apsrev4-1}

\end{document}